\documentclass[5p,authoryear,times]{elsarticle}
\usepackage{amssymb}
\usepackage{amsmath}
\usepackage[colorlinks]{hyperref}
\usepackage[utf8]{inputenc}
\usepackage[english]{babel}
\usepackage{amsmath,mathtools}
\DeclareMathOperator{\tr}{tr}

\usepackage[9pt]{extsizes}
\usepackage{cellspace}
\usepackage{setspace}

\journal{International Journal of Solids and Structures}

\begin{document}
\sloppy 
\allowdisplaybreaks

\begin{frontmatter}
 
\author[ioffe]{A.A.Semenov}

\author[ioffe]{Y.M. Beltukov\corref{cor1}} 
\ead{ybeltukov@gmail.com}

\address[ioffe]{Ioffe Institute, Politekhnicheskaya 26, 194021 St. Petersburg, Russia}

\cortext[cor1]{Corresponding author}

\title{\vspace{2.0cm}Nonlinear elastic moduli of composite materials with nonlinear spherical inclusions dispersed in a nonlinear matrix}

\begin{abstract}
A theory is developed for evaluation of nonlinear elastic moduli of composite materials with nonlinear inclusions dispersed in another nonlinear material (matrix). We elaborate a method aimed for determination of elastic parameters of a composite: its linear elastic moduli (second-order elastic constants) and nonlinear elastic moduli, which are known as the Murnaghan moduli (third-order elastic constants). We find an analytical form for the effective Murnaghan moduli of a composite with spherical inclusions. The effective moduli depend linearly on Murnaghan moduli of constituents. The results obtained have been verified through numerical modeling using the finite element method.
\end{abstract}


\end{frontmatter}
\section{Introduction}

Composite materials are increasingly used in everyday life (\cite{phil2014, ishikawa2018, toozandehjani2018}). In this paper, we consider composites from the point of view of their macroscopic mechanical properties. 
A number of techniques have been developed for strengthening of composite materials and for improvement of their elastic properties (\cite{fawaz2015}).
These techniques provide a variety of different types of composite materials which can be described by different theoretical models. Laminated composites can be considered as materials with piecewise homogeneous properties (\cite{nariboli1976, gula2017}). Composites containing various kinds of particles, usually small-size and evenly distributed, can be regarded in most of models as homogeneous materials (\cite{ColomboGiordano, giordano2017}). Fiber-reinforced composites and those with large-size particles  (\cite{mortazavian2015, giordano2013}) exhibit intermediate properties (\cite{zaferani2018}). In macro-mechanical models, all the above-mentioned composites except laminated ones can be considered as homogeneous materials with modified properties. 

The use of composite materials allows to achieve unique physical and mechanical properties. 
Under certain conditions, composites can provide better properties than those of the matrix material and inclusions separately (\cite{koratkar2005, uddin2008}).
Therefore, the determination and prediction of elastic properties of such composite materials are highly desirable for various applications.

In the first approximation, elastic properties of a material are governed by the linear elastic moduli, which characterize elastic stresses of a solid at small strains. Since linear moduli give a quadratic contribution to the elastic energy, they are also called the second-order moduli. Linear elastic properties of composite materials have been extensively studied both in theory and experiments and expressions for second-order moduli of elasticity have been derived (\cite{eshelby57, HalpinKardos}).

However, as strain increases, nonlinear elastic properties start to contribute to material behaviour. Nonlinear elasticity of a material is governed by the third-order elastic moduli. These nonlinear moduli are responsible, among other things, for the formation and propagation of nonlinear strain solitary waves (solitons for brevity) (\cite{samsonovbook}). 

The first mathematical description of such waves was made by Korteweg and de Vries (\cite{korteweg1895}). At the end of the last century, the theory developed by A.\,M.~Samsonov described the propagation of bulk solitons in solids (see \cite{samsonovbook} and references therein) and became a basis for first successful experiments on generation and observation of bulk strain solitons in nonlinearly elastic solid waveguides (\cite{Ostrovsky1988}).

The basic parameter governing soliton formation and propagation in a solid is a nonlinearity coefficient comprising a combination of linear and nonlinear elastic moduli of the material (\cite{samsonovbook}). Thus, for correct prediction and description of soliton formation and behaviour in a composite waveguide, one should know the effective nonlinear elastic moduli of the composite. The assessment of these values is thus in high demand. 

There are several models of materials that take into account nonlinear corrections. In these models both elastic (\cite{murnaghan1937}) and inelastic (\cite{mooney1940, cohen1993}) materials are analyzed. We will consider only elastic deformations, which means that after an applied deformation, the body returns to its original non-deformed state. Therefore, we will use the Murnaghan’s  theory where the potential energy density is expressed in the form:
\begin{equation} \label{eq:PotentialEnrgMurnaghan}
\Pi = \frac{\lambda + 2\mu}{2} \mathrm{I}_1^2 - 2\mu  \mathrm{I}_2 + \frac{l+2m}{3} \mathrm{I}_1^3 - 2m  \mathrm{I}_1  \mathrm{I}_2 + n  \mathrm{I}_3,
\end{equation}
where $I_1, I_2, I_3$ are invariants of the strain tensor $\hat{\varepsilon}$:
\begin{equation}
 \mathrm{I}_1 = \tr \hat{\varepsilon}; \quad  \mathrm{I}_2 = \frac12 \left( (\tr\hat{\varepsilon})^2 - \tr\hat{\varepsilon}^2 \right); \quad  \mathrm{I}_3 = \det \hat{\varepsilon},
\end{equation}
$l,m,n$ are Murnaghan (third-order) moduli and $\lambda$, $\mu$ are Lame (second-order) parameters.
Elastic moduli are coefficients in the power expansion of the potential energy density, and the set of these values defines elastic properties of the material. Third-order elastic moduli show the deviation from the linear elasticity of solids. 

To date, several experimental studies have been carried out on measurements of the third-order elastic moduli. However, values measured for the same material can vary considerably in different experiments (\cite{renaud2016, winkler2004}). Moreover, measurement error is often huge and  can reach 100\% (\cite{LangGupta2011, hughes1953, korobov2013}). Therefore, a theoretical approach is needed for an accurate evaluation of nonlinear elastic moduli of composites.

There have been several theoretical studies performed on the subject. The Eshelby’s theory (\cite{eshelby57}) for a single ellipsoidal inclusion was generalized to the case of nonlinear elastic properties of inclusion  (\cite{sevostianov2001,tsvelodub2000,tsvelodub2004}). Macroscopic elastic properties of composites with linear matrix and nonlinear inclusions was obtained by S. Giordano and L. Colombo  (\cite{giordano2009, ColomboGiordano, giordano2017}).
In practice, stiffer inclusions are often introduced into a softer matrix, for example, to increase the stiffness of the resulting nanocomposite. Therefore, during deformation of such a composite material, the matrix undergoes greater deformation than inclusions. Accordingly, in this case, the nonlinear properties of the matrix play a more essential role than the nonlinear properties of the inclusions.

In this paper, we consider the general case when both the inclusions and the matrix possess nonlinear elastic moduli with an arbitrary ratio between the stiffness of the inclusions and the matrix. We also assume the applicability of the nonlinear theory of elasticity. In other words, we assume that the inclusions are of a sufficiently large size, at which the molecular structure of the substance does not play a significant role.

The paper is organized as follows. In Section~\ref{sec:formulation} the nonlinear elasticity problem for the composite is formulated. For a given linear and nonlinear elastic properties of composite constituents, we provide a procedure to find effective values of macroscopic moduli. In Section~\ref{sec:reduction} linear iterations are used to solve the nonlinear problem. In Section~\ref{sec:eshelby} the Eshelby's theory is applied to the linearized problem. In Section~\ref{sec:moduli} we find the effective linear and nonlinear moduli of a composite with spherical inclusions. In Section \ref{sec:num} the obtained formulas are verified using the finite element method. Finally, in Section \ref{sec:conclusion} we discuss the results obtained.

\section{Formulation of the problem}
\label{sec:formulation}

Let us consider a certain volume $\Omega$, where a composite consisting of a matrix and inclusions is located. To determine effective elastic moduli of the composite, let us apply some external deformation to it. The following strain tensor can be written for a point $\vec{x}$ in equilibrium:
\begin{equation}
    \varepsilon_{i\!j}(\vec{x}) = \frac{1}{2}\Big(u_{i\!j}(\vec{x})+u_{ji}(\vec{x})+u_{ki}(\vec{x})u_{kj}(\vec{x})\Big),  \label{eq:def}
\end{equation}
where
\begin{equation}
    u_{i\!j}(\vec{x}) = \frac{\partial u_i(\vec{x})}{\partial x_j}
\end{equation}
is the distortion tensor (strain gradient).

Since we are interested in nonlinear properties of the composite, the nonlinear term in the definition of the strain tensor (\ref{eq:def}) cannot be considered small. We assume that deformations are finite, and cannot be considered small enough to disregard the nonlinear term in (\ref{eq:def}). We assume also that deformations are elastic, which means that after unloading, the body returns to its original state.

At each point $\vec{x}$, the potential energy density $\Pi(\vec{x})$ depends on the strain tensor $\varepsilon_{i\!j}(\vec{x})$ and on the values of elastic moduli of the material at that point. In general, this dependence can be represented as an expansion in powers of $u_{i\!j}(\vec{x})$:
\begin{equation}
    \Pi(\vec{x}) = \frac{1}{2}C_{i\!jkl}(\vec{x})u_{i\!j}(\vec{x})u_{kl}(\vec{x}) + \frac{1}{3}N_{i\!jklmn}(\vec{x})u_{i\!j}(\vec{x})u_{kl}(\vec{x})u_{mn}(\vec{x}) + O(u^4).
\label{eq:Pi(x)}
\end{equation}
For homogeneous and isotropic body, linear and nonlinear stiffness tensors have the following form:
\begin{gather}
    C_{i\!jkl} = \lambda  \delta_{i\!j} \delta_{kl}+\mu 
    \left(\delta_{il} \delta_{jk}+\delta
    _{ik} \delta_{jl}\right),   \label{C} \\
    \begin{multlined}
        N_{i\!jkhts} = 
        \frac{\lambda}{2} \left(\delta_{hk}
        \delta_{it} \delta_{js}+\delta_{hs}
        \delta_{i\!j} \delta_{kt}+\delta_{hj}
        \delta_{ik} \delta_{st}\right)+
        \frac{\mu}{2} \Big(\delta_{hs} \delta_{it} \delta_{jk}\\
        + \delta_{ht} \delta_{ik} \delta_{js} 
        + \delta_{hs} \delta_{ik} \delta_{jt}
        + \delta_{hj} \delta_{it} \delta_{ks}
        + \delta_{hj} \delta_{is} \delta_{kt}
        + \delta_{hi} \delta_{js} \delta_{kt}\Big)\\
        + l \delta_{hk} \delta_{i\!j} \delta_{st}
        + 
        \frac{m}{2}\Big(\delta_{hj} \delta_{ik} \delta_{st}
        + \delta_{hi} \delta_{jk} \delta_{st}
        + \delta_{hk} \delta_{it} \delta_{js}
        + \delta_{hk} \delta_{is} \delta_{jt} \\
        + \delta_{ht} \delta_{i\!j} \delta_{ks}
        + \delta_{hs} \delta_{i\!j} \delta_{kt}
        - 2\delta_{hk} \delta_{i\!j} \delta_{st}\Big)
        \\
        +\frac{n}{8}\Big(
        \delta_{ht} \delta_{is} \delta_{jk}
        +\delta_{hs} \delta_{it} \delta_{jk}
        -2 \delta_{hi} \delta_{jk} \delta_{st}
        +\delta_{ht} \delta_{ik} \delta_{js}
        -2 \delta_{hk} \delta_{it} \delta_{js}
        \\
        +\delta_{hs} \delta_{ik} \delta_{jt}
        -2 \delta_{hk} \delta_{is} \delta_{jt} 
        -2 \delta_{ht} \delta_{i\!j}\delta_{ks}
        +\delta_{hj} \delta_{it} \delta_{ks}
        -2 \delta_{hj} \delta_{ik} \delta_{st}
        \\
        +\delta_{hj} \delta_{is} \delta_{kt}
        -2 \delta_{hs} \delta_{i\!j} \delta_{kt}
        +\delta_{hi} \delta_{js} \delta_{kt}
        +4 \delta_{hk} \delta_{i\!j} \delta_{st}
        +\delta_{hi} \delta_{jt} \delta_{ks}
        \Big),
        \label{N}
    \end{multlined}
\end{gather}
where $l,m,n$ are Murnaghan moduli and $\lambda$, $\mu$ are Lame parameters. Further, we will mainly use a pair of linear elastic moduli $K$, $\mu$, since the resulting equation has a simpler form in these notations. Here $K$ is the bulk elastic modulus, which is expressed through $\lambda$, $\mu$ in the form:
\begin{equation}
	K = \lambda + \frac{2 \mu}{3}.
\end{equation}

In a general case, the elastic properties of the matrix and inclusions are different. For both materials we have different sets of linear and non-linear elastic moduli: $K_0, \mu_0, l_0, m_0, n_0$ for the matrix and $K_1, \mu_1, l_1, m_1, n_1$ for the inclusions. Thus, volumes $\omega_{i}$ containing inclusions are characterized by elastic moduli tensors $\hat{C}^1$ and $\hat{N}^1$. The remaining volume $\Omega \setminus (\omega_1 \cup \omega_2 \cup ... )$ containing matrix material is characterized by elastic moduli tensors $\hat{C}^0$ and $\hat{N}^0$. To obtain effective macroscopic elastic moduli of the composite, we need to average the elastic moduli over the composite volume $\Omega$ in a special way. 

\begin{figure}[t]
    \centering
    \includegraphics[width=6cm]{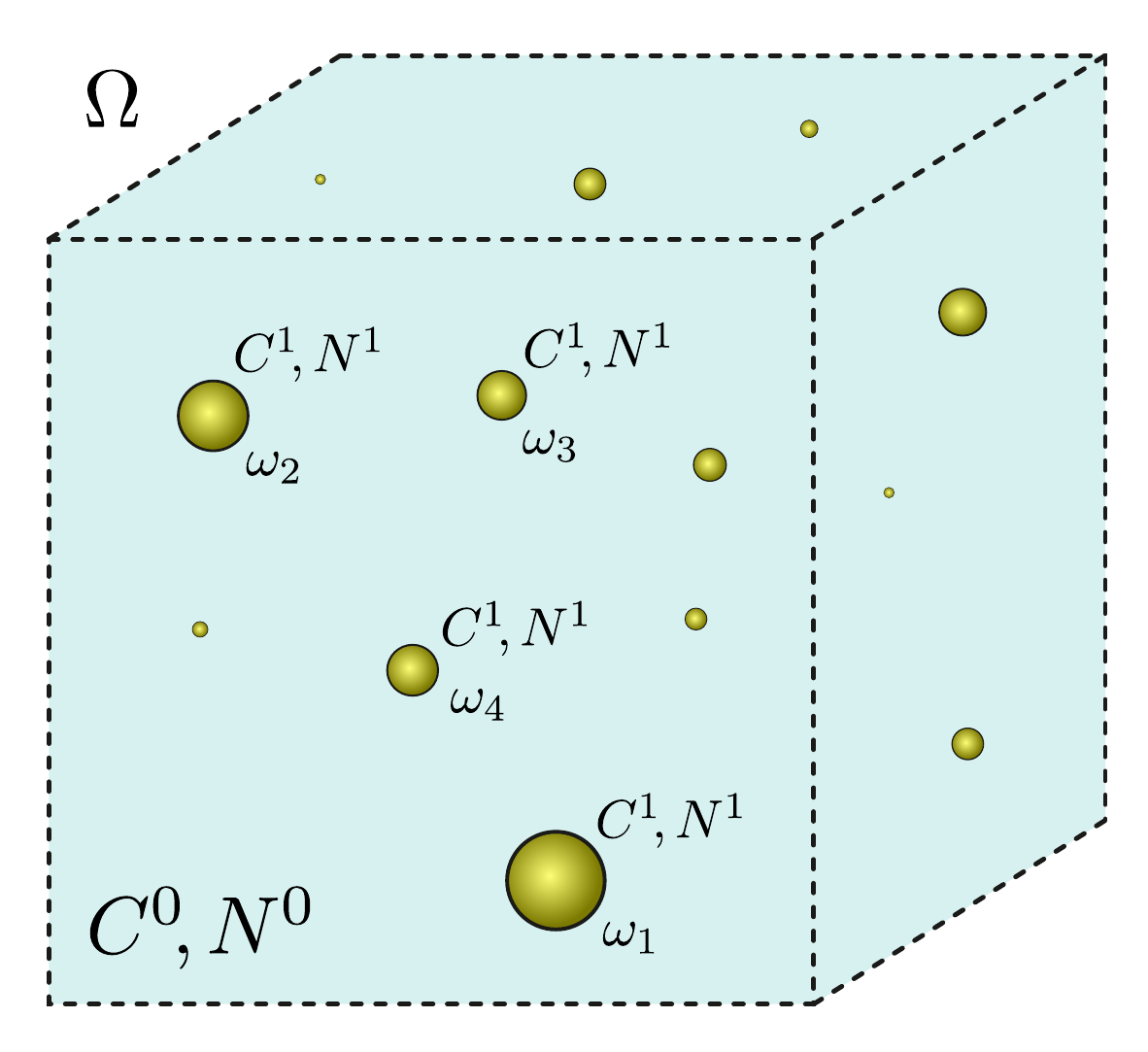}
    \caption{(Color online) Schematic of the composite of volume $\Omega$ containing spherical inclusions of volume $\omega_{i}$. $C^0, N^0$ are elastic moduli tensors of the matrix. $C^1, N^1$ are those of the inclusions.}
    \label{fig:composite}
\end{figure}

To apply a deformation to the composite, we assume that the boundary conditions have a form
\begin{equation}
    u_i(\vec{x})\big|_{\partial\Omega} = u^0_{i\!j} x_j,  \label{eq:bc}
\end{equation}
where the boundary deformation $u^0_{i\!j}$ is given and does not depend on $\vec{x}$. In the case of a homogeneous medium in equilibrium, we would have a homogeneous deformation, which is described by a displacement $u_i(\vec{x}) = u^0_{i\!j}x_j$. However, for a composite or any other inhomogeneous medium, the displacement $u_i(\vec{x})$ is not a linear function of $\vec{x}$ coordinate. Consider a volume-average distortion tensor
\begin{equation}
    \overline{u}_{i\!j} = \frac{1}{V}\int_\Omega u_{i\!j}(\vec{x})d\vec{x}.
\end{equation}
Note that the equality $\overline{u}_{i\!j}=u^0_{i\!j}$ is satisfied for a nonlinear elastic medium as well. Indeed,
\begin{multline}
    \overline{u}_{i\!j} = \frac{1}{V}\int_\Omega u_{i\!j}(\vec{x})d\vec{x} = \frac{1}{V}\int_{\partial\Omega} u_i(\vec{x}) n_j dS = \\
    \frac{1}{V}\int_{\partial\Omega} u^0_{ik} x_k n_j dS = \frac{1}{V}\int_{\Omega} u^0_{ik} x_{kj} d\vec{x} = u^0_{i\!j}.
\end{multline}

For each applied deformation (defined by $\overline{u}_{i\!j}=u^0_{i\!j}$), the average elastic energy of the composite is
\begin{equation}
    \overline{\Pi} = \frac{1}{V}\int\Pi(\vec{x})d\vec{x}.  \label{eq:avPi}
\end{equation}
It can be represented as an expansion in powers of $\overline{u}_{i\!j}$:
\begin{equation}
    \overline{\Pi} = \frac{1}{2}C^{\rm eff}_{i\!jkl} \overline{u}_{i\!j} \overline{u}_{kl} + \frac{1}{3}N^{\rm eff}_{i\!jklmn} \overline{u}_{i\!j} \overline{u}_{kl} \overline{u}_{mn} + O\left(\overline{u}^{\,4}\right),
    \label{eq:eff}
\end{equation}
which defines the effective elastic moduli tensors $\hat{C}^{\rm eff}$ and $\hat{N}^{\rm eff}$. Since the composite is homogeneous and isotropic at the macroscopic scale, the effective elastic moduli tensors $\hat{C}^{\rm eff}$ and $\hat{N}^{\rm eff}$ have the same general form as $\hat{C}$ and $\hat{N}$ (Eqs.~(\ref{C}) and (\ref{N})) with effective elastic moduli $\lambda_{\rm eff}$, $\mu_{\rm eff}$, $l_{\rm eff}$, $m_{\rm eff}$, $n_{\rm eff}$.

Equation (\ref{eq:eff}) can be solved both analytically and numerically. In Sections \ref{sec:reduction}--\ref{sec:moduli} we perform an analytical averaging of the elastic energy and find an analytical expression for the effective elastic moduli. In Section \ref{sec:num} we find numerical values of effective moduli using the finite element method.

\section{Reduction to a linear problem and assumptions}
\label{sec:reduction}

Let us solve the problem iteratively by the method of successive approximations. The distortion tensor $u_{i\!j}(\vec{x})$ can be written as
\begin{equation}
    u_{i\!j}(\vec{x}) = u_{i\!j}^L(\vec{x})+u_{i\!j}^{N}(\vec{x}),
    \label{eq:L+N}
\end{equation}
where $u_{i\!j}^L(\vec{x})$ is the solution to the linear problem and $u_{i\!j}^{N}(\vec{x}) \ll u_{i\!j}^L(\vec{x})$ is a small nonlinear contribution.
Then the boundary conditions take the form:
\begin{gather}
    u_i^L\big|_{\partial\Omega} = u_{i\!j}^0 x_j, \\
    u_i^N\big|_{\partial\Omega} = 0.
\end{gather}
Using Eqs.~(\ref{eq:Pi(x)}) and (\ref{eq:L+N}) and taking into account that $u_{i\!j}^{N}(\vec{x}) \ll u_{i\!j}^{L}(\vec{x})$, the average density of potential energy can be written in the form:
\begin{multline}
    \overline{\Pi} = \frac{1}{V}\int\Bigl( \frac{1}{2} C_{i\!jkl}(\vec{x})u_{i\!j}^L(\vec{x})u_{kl}^L(\vec{x}) +C_{i\!jkl}(\vec{x})u_{i\!j}^L(\vec{x})u_{kl}^N(\vec{x}) \\
    +\frac{1}{3} N_{i\!jklmn}(\vec{x})u_{i\!j}^L(\vec{x})u_{kl}^L(\vec{x})u_{mn}^L(\vec{x})\Bigl)d\vec{x}.
\end{multline}
Integrating the second term in parts and applying the Gauss theorem (\cite{giordano2013}), we note that
\begin{equation}
    \overline{\Pi} = \frac{1}{V}\int\Bigl( \frac{1}{2}C_{i\!jkl}(\vec{x})u_{i\!j}^L(\vec{x})u_{kl}^L(\vec{x})
    +\frac{1}{3} N_{i\!jklmn}(\vec{x})u_{i\!j}^L(\vec{x})u_{kl}^L(\vec{x})u_{mn}^L(\vec{x})\Bigl)d\vec{x}.
\end{equation}
Thus, in order to find the average density of potential energy with the required accuracy, it is necessary to calculate the linear deformation only, i.e. to solve a linear problem.

We assume that all inclusions in the composite are the same and are located far enough from each other. Therefore, interaction between inclusions can be neglected. Then the density of the elastic energy far from inclusions can be expressed as:
\begin{equation}
    \Pi^\infty = \frac{1}{2}C^0_{i\!jkl} u^\infty_{i\!j} u^\infty_{kl} + \frac{1}{3}N^0_{i\!jklmn} u^\infty_{i\!j} u^\infty_{kl} u^\infty_{mn},
    \label{eq:Pi_infty}
\end{equation}
where $u^\infty_{i\!j} \ne \overline{u}_{i\!j}$ is the distortion tensor far from inclusions. Therefore, we can write the average density of potential energy as
\begin{equation}
    \overline{\Pi} = \frac{1}{V}\int_\Omega\Pi(\vec{x})d\vec{x} = \Pi^\infty+\frac{1}{V}\int_\Omega\Bigl(\Pi(\vec{x})-\Pi^\infty\Bigr)d\vec{x}.
    \label{eq:Pi_average1}
\end{equation}
One can see that ${\Pi(\vec{x})-\Pi^\infty}$ goes to zero far from inclusions. Therefore, the integral in Eq.~(\ref{eq:Pi_average1}) can be calculated using the contribution from each inclusion independently:
\begin{equation}
    \overline{\Pi} = \Pi^\infty+\Bigl<\Pi^1(\vec{x})-\Pi^\infty\Bigr>_{\vec{x}},
    \label{eq:Pi_average}
\end{equation}
where $\Pi^1(\vec{x})$ is the contribution to the potential energy density from one inclusion:
\begin{equation}
    \Pi^1(\vec{x}) = \frac{1}{2}C_{i\!jkl}(\vec{x}) u^1_{i\!j}(\vec{x}) u^1_{kl}(\vec{x}) + \frac{1}{3}N_{i\!jklmn}(\vec{x}) u^1_{i\!j}(\vec{x}) u^1_{kl}(\vec{x}) u^1_{mn}(\vec{x}),
    \label{eq:Pi_1}
\end{equation}
with $u^1_{i\!j}(\vec{x})$ being the distortion around the inclusion $\omega_1$ and inside it.  The averaging in Eq. (\ref{eq:Pi_average}) is performed using the following homogenization procedure:
\begin{equation}
    \Bigl< f(\vec{x})\Bigr>_{\vec{x}} = 
    \frac{N}{V}\Biggl\{\int_{\vec{x}\in \omega_1}f(\vec{x})d\vec{x} +  (1-c)\int_{\vec{x}\notin \omega_1}f(\vec{x})d\vec{x} \Biggr\},   \label{eq:averaging}
\end{equation}
where $N$ is the number of inclusions and $c$ is the volume fraction of inclusions. The first integral in Eq. (\ref{eq:averaging}) is performed over an inclusion volume $\omega_1$. The second integral is performed over the surrounding matrix. The coefficient $(1-c)$ takes into account that the surrounding matrix is partially occupied by other inclusions. It is the simplest procedure, which takes into account the presence of neighbor inclusions. As we will see in Section~\ref{sec:num}, it gives accurate effective moduli in a wide range of the volume fraction $c$.

Similarly, the average distortion tensor $\overline{u}_{i\!j}$ can be written in the form
\begin{equation}
    \overline{u}_{i\!j} = \frac{1}{V}\int_\Omega u_{i\!j}(\vec{x})d\vec{x} = u_{i\!j}^\infty+\Bigl<u_{i\!j}^1(\vec{x})-u_{i\!j}^\infty\Bigr>_{\vec{x}}.
    \label{eq:averageu}
\end{equation}
Thus, we can consider inclusions separately from each other and find $u_{i\!j}^1(\vec{x})$ and $\Pi^1(\vec{x})$. Then, the effective elastic moduli can be found using Eqs. (\ref{eq:Pi_average}), (\ref{eq:averageu}) and (\ref{eq:eff}).

\section{Eshelby’s theory}
\label{sec:eshelby}

A solution to the linear problem is known from the theory developed by Eshelby (\cite{eshelby57, ColomboGiordano}).
The strain field of a matrix with a single inclusion $\omega_1$ can be expressed using the ``eigen-strain'' parameter $u_{i\!j}^*$:
\begin{equation}
    u_{i\!j}^1(\vec{x})  = L_{i\!jkl}(\vec{x}) u_{kl}^*,
    \label{eq:u_1_to_u*}
\end{equation}
where 
\begin{equation}
    L_{i\!jkl}(\vec{x}) = S_{i\!jkl}(\vec{x}) + L_{i\!jkl}^\infty.
    \label{eq:L(x)}
\end{equation}
Tensor $\hat{L}^\infty$ is defined as
\begin{equation}
    \hat{L}^\infty = \bigl(\hat{I} - \bigl(\hat{C}^{0}\bigr)^{-1} \hat{C}^{1}\bigr)^{-1} - \hat{S}^0.
    \label{eq:Linfty}
\end{equation}
The Eshelby’s tensor is defined as (\cite{eshelby57})
\begin{multline}
    S_{i\!jkl}(\vec{x}) = \frac{1}{8\pi(1-\nu_0)}\frac{\partial^4\Psi}{\partial x_i\partial x_j\partial x_k\partial x_l}-\frac{\nu_0}{1-\nu_0}\frac{\delta_{kl}}{4\pi}\frac{\partial^2\Phi}{\partial x_i\partial x_j} \\
    - \frac{1}{8 \pi }\left(
     \delta_{il}\frac{\partial^2\Phi}{\partial x_j\partial x_k}
    +\delta_{ik}\frac{\partial^2\Phi}{\partial x_j\partial x_l}
    +\delta_{jl}\frac{\partial^2\Phi}{\partial x_i\partial x_k}
    +\delta_{jk}\frac{\partial^2\Phi}{\partial x_i\partial x_l}\right),
    \label{eq:Eshelby}
\end{multline}
where harmonic $\Phi(\vec{x})$ and bi-harmonic $\Psi(\vec{x})$ potentials have the form
\begin{gather}
    \Phi(\vec{x}) = \int_{\omega_1} \frac{1}{|\vec{x}-\vec{x}_1|}d\vec{x}_1, \\
    \Psi(\vec{x}) = \int_{\omega_1} |\vec{x}-\vec{x}_1| d\vec{x}_1.
\end{gather}
For elliptic inclusions, the Eshelby’s tensor is constant inside inclusions: ${\hat{S}(\vec{x}) = \hat{S}^0}$ for $\vec{x} \in \omega_1$. As a result, the potential energy density from a single particle $\Pi^1(\vec{x})$ has the form
\begin{multline}
    \Pi^1(\vec{x}) = \frac{1}{2} C_{i\!jkl}(\vec{x})L_{i\!ji'\!\!j'}(\vec{x})L_{klk'\!l'}(\vec{x})u_{i'\!\!j'}^*u_{k'\!l'}^* \\
    +\frac{1}{3}N_{i\!jklmn}(\vec{x})L_{i\!ji'\!\!j'}(\vec{x})L_{klk'\!l'}(\vec{x})L_{mnm'\!n'}(\vec{x})u_{i'\!\!j'}^*u_{k'\!l'}^*u_{m'\!n'}^*.
    \label{eq:Pi_1_to_u*}
\end{multline}
The Eshelby’s theory also allows to find the distortion tensor far from the inclusion $\hat{u}^\infty$:
\begin{equation}
    u_{i\!j}^\infty = L^\infty_{i\!jkl} u_{kl}^*.
    \label{eq:uinfty_to_u*}
\end{equation}
Thus, the density of potential energy far from inclusion has a form
\begin{equation}
\Pi^\infty = \frac{1}{2} C^0_{i\!jkl}L^\infty_{i\!ji'\!\!j'}L^\infty_{klk'\!l'}u_{i'\!\!j'}^*u_{k'\!l'}^*+ \frac{1}{3}N^0_{i\!jklmn}L^\infty_{i\!ji'\!\!j'}L^\infty_{klk'\!l'}L^\infty_{mnm'\!n'}u_{i'\!\!j'}^*u_{k'\!l'}^*u_{m'\!n'}^*.
\label{eq:Pi_infty_to_u*}
\end{equation}

\section{Effective elastic moduli of a composite with spherical inclusions}
\label{sec:moduli}

Substituting  (\ref{eq:Pi_1_to_u*}) and (\ref{eq:Pi_infty_to_u*}) into (\ref{eq:Pi_average}), we obtain the average deformation of the composite expressed through the eigen-strain parameter  $\hat{u}^*$:
\begin{equation}
    \overline{\Pi} = \frac{1}{2}C_{i'\!\!j'\!k'\!l'}^*u_{i'\!\!j'}^*u_{k'\!l'}^*+\frac{1}{3}N_{i'\!\!j'\!k'\!l'\!m'\!n'}^* u_{i'\!\!j'}^*u_{k'\!l'}^*u_{m'\!n'}^*,
    \label{eq:average_Pi_to_u*}
\end{equation}
where
\begin{gather}
    C_{i'\!\!j'\!k'\!l'}^* = 
    C_{i\!jkl}^0 L_{i\!ji'\!\!j'}^\infty L_{klk'\!l'}^\infty 
    +\Bigl< C_{i\!jkl}(\vec{x}) L_{i\!ji'\!\!j'}(\vec{x}) L_{klk'\!l'}(\vec{x}) - C_{i\!jkl}^0 L_{i\!ji'\!\!j'}^\infty L_{klk'\!l'}^\infty \Bigr>_{\vec{x}}, \label{eq:C*} \\
    \begin{multlined}
        N_{i'\!\!j'\!k'\!l'\!m'\!n'}^* = 
        N_{i\!jklmn}^0 L_{i\!ji'\!\!j'}^\infty L_{klk'\!l'}^\infty L_{mnm'\!n'}^\infty\\
        +\Bigl< N_{i\!jklmn}(\vec{x}) L_{i\!ji'\!\!j'}(\vec{x}) L_{klk'\!l'}(\vec{x}) L_{mnm'\!n'}(\vec{x}) - N_{i\!jklmn}^0 L_{i\!ji'\!\!j'}^\infty L_{klk'\!l'}^\infty L_{mnm'\!n'}^\infty \Bigr>_{\vec{x}}. 
    \end{multlined} 
    \label{eq:N*}
\end{gather} 
On the other hand, by substitution of Eqs.~(\ref{eq:uinfty_to_u*}) and (\ref{eq:u_1_to_u*}) into Eq.~(\ref{eq:averageu}) we can also obtain the average deformation of the composite expressed through the eigen-strain parameter  $\hat{u}^*$:
\begin{equation}
\overline{u}_{i\!j} = \hat{L}_{i\!ji'\!\!j'}^* \hat{u}_{i'\!\!j'}^*
\label{eq:average_u_to_u*}
\end{equation}
where
\begin{equation}
    L_{i\!ji'\!\!j'}^* =  L_{i\!ji'\!\!j'}^\infty + \Bigl<L_{i\!ji'\!\!j'}(\vec{x})-L_{i\!ji'\!\!j'}^\infty \Bigr>_{\vec{x}}.
    \label{eq:L*}
\end{equation}
Then, according to  (\ref{eq:eff}), we can express the average density of potential energy by using the tensor $\hat{L}^*$ and the eigen-strain parameter $\hat{u}^*$:
\begin{equation}
\overline{\Pi} = \frac{1}{2} C^{\rm eff}_{i\!jkl}L^*_{i\!ji'\!\!j'}L^*_{klk'\!l'}u_{i'\!\!j'}^*u_{k'\!l'}^* +\frac{1}{3}N^{\rm eff}_{i\!jklmn}L^*_{i\!ji'\!\!j'}L^*_{klk'\!l'}L^*_{mnm'\!n'}u_{i'\!\!j'}^*u_{k'\!l'}^*u_{m'\!n'}^*.
\label{eq:average+eff_Pi_to_u*}
\end{equation}
Comparing equations (\ref{eq:average_Pi_to_u*}) and (\ref{eq:average+eff_Pi_to_u*}) for the average density of potential energy of the composite, we obtain two systems of algebraic equations for elements of the effective linear and nonlinear stiffness tensors $\hat{C}^{\rm eff}$ and $\hat{N}^{\rm eff}$. The system of equations for the linear moduli has a form:
\begin{equation}
C_{i\!jkl}^{\rm eff} L_{i\!ji'\!\!j'}^*L_{klk'\!l'}^* = C_{i'\!\!j'\!k'\!l'}^*.
\label{seq:Ceff}
\end{equation}
The system of equations for the nonlinear moduli has a similar form:
\begin{equation}
N_{i\!jklmn}^{\rm eff} L_{i\!ji'\!\!j'}^*L_{klk'\!l'}^*L_{mnm'\!n'}^* = N_{i'\!\!j'\!k'\!l'\!m'\!n'}^*.
\label{seq:Neff}
\end{equation}

The harmonic and bi-harmonic potentials have a simple form for spherical inclusions of radius~$R$:  
\begin{gather}
    \Phi(\vec{r}) = 
    \begin{cases}
    \dfrac{2\pi}{3} (3 R^2 - r^2),& r<R,\\
    \dfrac{4\pi R^3}{3r},& r>R.
    \end{cases} \\
    \Psi(\vec{r}) = 
    \begin{cases}
    \dfrac{\pi r^2}{15} (10 R^2 - r^2),& r<R,\\
    \dfrac{\pi  R^3}{15r} \left(4 R^2+20r^2-15 Rr\right),& r>R.
    \end{cases}
\end{gather}
Using the computer algebra system, we make the homogenization procedure in Eqs.~(\ref{eq:C*}), (\ref{eq:N*}), (\ref{eq:L*}) and solve systems of equations (\ref{seq:Ceff}) and (\ref{seq:Neff}). As a result, we get effective linear and nonlinear moduli of a composite with spherical inclusions. The linear effective elastic moduli can be expressed in terms of elastic moduli of the matrix and volume fraction of inclusions $c = 4\pi N R^3 / 3 V$:
\begin{gather}
    K_{\rm eff} = K_0 + \frac{c (K_1-K_0)(3K_0+4\mu_0)}{3 K_1+4 \mu_0 - 3c(K_1 - K_0)},  \label{eq:K_eff}  
    \\
    \mu_{\rm eff} = \mu_0 + \frac{5c\mu_0(\mu_1-\mu_0)(3K_0+4\mu_0)}{\mu_0((6c+9)K_0+4(3c+2)\mu_0)+6(1-c)\mu_1\left(K_0+2\mu_0\right)}.  \label{eq:mu_eff}
\end{gather}
This result coincides with Eqs.~(15) and (16) in (\cite{giordano2008}).
Nonlinear effective elastic moduli of the composite take the form:
\begin{equation}
    \begin{pmatrix} 
        l_{\rm eff}\\m_{\rm eff}\\n_{\rm eff}
    \end{pmatrix}
    = 
    \begin{pmatrix} 
        l_0\\m_0\\n_0
    \end{pmatrix}
    + c  \mathbf{P}^{0} \cdot 
    \begin{pmatrix} 
        l_0\\m_0\\n_0
    \end{pmatrix}
    + c \mathbf{P}^{1} \cdot 
    \begin{pmatrix} 
        l_1\\m_1\\n_1
    \end{pmatrix}
    + c
    \begin{pmatrix} 
        l_g\\m_g\\n_g
    \end{pmatrix}.
    \label{eq:final}
\end{equation}
Here $\mathbf{P}^{0}$ and $\mathbf{P}^{1}$ are $3\times 3$ matrices, which depend on the linear moduli of the matrix $K_0, \mu_0$ and those of inclusions  $K_1, \mu_1$. In general, these matrices also depend on the volume fraction $c$. Nonlinear moduli $l_g$, $m_g$, $n_g$ provide the ``geometric'' nonlinear contribution which exists even if all Murnaghan moduli of the matrix and inclusions are zero. They also depend only on linear moduli $K_0, \mu_0, K_1, \mu_1$ and the volume fraction $c$. The analytical expressions for $\mathbf{P}^{0}$, $\mathbf{P}^{1}$, $l_g$, $m_g$, $n_g$ are presented in \ref{app}.

The same procedure can be applied to obtain effective nonlinear moduli for composites with ellipsoidal inclusions (aligned or randomly oriented). However, it involves more complicated integrals in the homogenization procedure (\ref{eq:averaging}) in Eqs.~(\ref{eq:C*}), (\ref{eq:N*}), (\ref{eq:L*}). For randomly oriented ellipsoids we expect the same form of the effective moduli (\ref{eq:final}) with coefficients depending also on the ellipsoidal aspect ratio. The proposed method can also be used to take into account spatial correlations of the inclusion positions. For example, one can consider so-called ellipsoidal microstructure with ellipsoidal symmetry of spatial correlations between the inclusions positions (\cite{giordano2017}). It requires an accurate modification of the homogenization procedure (\ref{eq:averaging}), which should be the subject of further work.

\section{Numerical verification}
\label{sec:num}

In order to verify the obtained equations, we performed numerical calculations using the finite element method. Note that to obtain the most reliable results, it is important to verify a general case of the final equation (\ref{eq:final}). Observe that in the case of relatively rigid inclusions (e.g. silica inclusions in a polymeric matrix) equations can be additionally simplified. Therefore, we have chosen a pair of materials, with elastic moduli of the same order of magnitude. We use two materials with known linear and nonlinear moduli with different Poisson ratios: polycarbonate matrix and polystyrene inclusions. Elastic moduli of both materials are given in Table~\ref{tab:materials}. We emphasize that the final result (\ref{eq:final}) can be applied for any pair of materials.

\begin{table}[t]
    \caption{Elastic moduli of polycarbonate and polystyrene (GPa)~(\cite{hughes1953, kruger1991, dreiden2011}).}
    \label{tab:materials}
    \centering
    \begin{tabular}{@{\hskip3pt}Sc@{\hskip10pt}Sc@{\hskip10pt}Sc@{\hskip10pt}Sc@{\hskip10pt}Sc@{\hskip10pt}Sc@{\hskip3pt}}
                \hline
                         &   $K$  &$\mu$&   $l$  &   $m$&   $n$\\
                \hline
        Polycarbonate   &   3.93 & 0.84  & $-50.0$ & $-12.2$ & $-32.0$\\
        Polystyrene     &   4.20 & 1.50  & $-18.9$ & $-13.3$ & $-10.0$\\
                \hline
    \end{tabular}
\end{table}

\begin{figure}[t]
    \centering
    \includegraphics[width=0.5\textwidth]{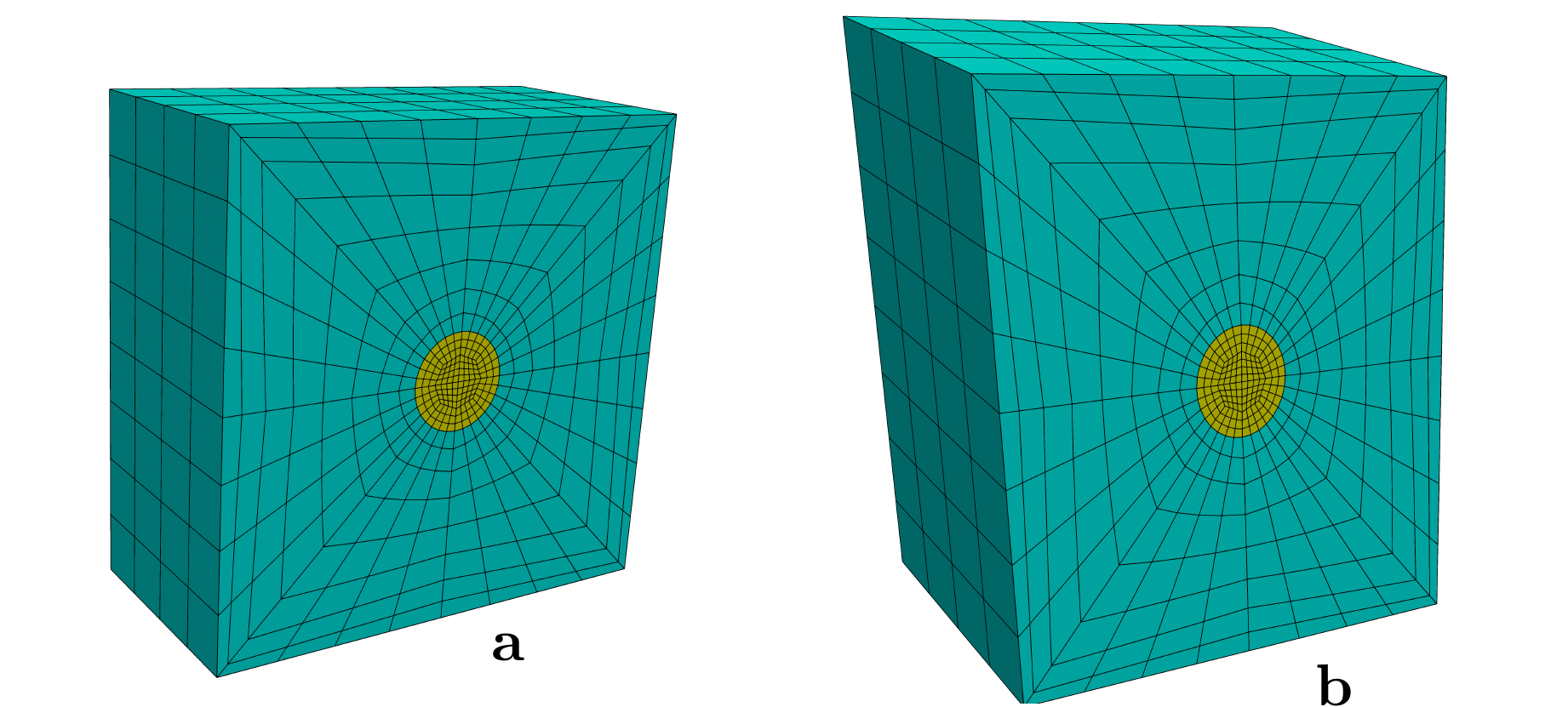}
    \caption{(Color online) A composite with the volume fraction of inclusion $c=0.4\%$ used in the finite element method: the spherical inclusion (yellow) inside the matrix (cyan). Only one half of the composite is shown to visualize the inclusion. a)~Undeformed composite. b)~Composite under deformation. The mesh has $N_0 = 4736$ hexahedral finite elements.}
    \label{fig:fem}
\end{figure}

Let us consider a cubic sample $L\times L\times L$ with a small inclusion of a radius $R$ located in the center (Fig.~\ref{fig:fem}). We apply a number of different trial distortions $u_{\smash{ij}}^0$  to obtain all elements of linear and nonlinear stiffness tensors $\hat{C}^{\rm eff}$ and $\hat{N}^{\rm eff}$ ($3^4 + 3^6 = 810$ in total). In order to reduce the influence of boundaries, we use periodic boundary conditions. In this case, we find the displacement in the form
\begin{equation}
    u_i(\vec{x}) = u^0_{i\!j} x_j + \tilde{u}_i(\vec{x})
\end{equation}
with $\tilde{u}_i(\vec{x})$ being the periodic function with the period $L$ in $x,y,z$ directions. The sample with one inclusion and periodic boundary conditions effectively acts as an infinite composite with inclusions placed at nodes of a simple cubic lattice. 

We use trial distortions of the form $u^0_{\smash{ij}} = \epsilon k_{\smash{ij}}$ where $\epsilon=10^{\smash{-5}}$ and $k_{\smash{ij}}$ are integer numbers, satisfying the condition $\sum_{ij} k_{\smash{ij}}^{\smash{2}} \le 2$. There are 163 trial distortions in total. For each distortion, we calculate all 9 elements of the average stress tensor $\hat{\overline{\sigma}}$, which have the following relation with the trial distortion $\hat{u}^0$:
\begin{equation}
    \overline{\sigma}_{ij} = C^{\rm eff}_{i\!jkl} u^0_{kl} + N^{\rm eff}_{i\!jklmn} u^0_{kl} u^0_{mn}.
\end{equation}
As a result, we obtain $9\cdot163 = 1467$ equations to find 810 unknown elements of $\hat{C}^{\rm eff}$ and $\hat{N}^{\rm eff}$. Since the number of equations is bigger than the number of unknown parameters, we use the least squares method to find the best solution. After the solution was found, we have checked that all equations were valid with a high enough precision. We deliberately do not rely on various symmetry properties of $\hat{C}^{\rm eff}$ and $\hat{N}^{\rm eff}$ to additionally verify the computations.

After all elements of the stiffness tensors $C^{\rm eff}$ and $N^{\rm eff}$ have been obtained, we could find the effective elastic moduli of the composite, $\lambda_{\rm eff}, \mu_{\rm eff}, l_{\rm eff}, m_{\rm eff}, n_{\rm eff}$, using Eqs.~(\ref{C}) and (\ref{N}). Once again, we used the least squares method to find the best values for the effective elastic moduli (5 values from 810 equations). For small-size inclusions the sample is effectively isotropic. For moderate values of the radius $R$, there is a small cubic anisotropy in the macroscopic behavior of the studied composite. However, the least squares method effectively takes the isotropic component of effective stiffness tensors.

We used Gmsh 2.10.1 (\cite{gmsh}) to prepare the hexagonal mesh (Fig.~\ref{fig:fem}). The geometry of the mesh was chosen to maximize the precision of the results for given computation time. Then we used FEniCS 2018.2.0 to apply the finite element method~(\cite{fenics}). It is open-source software, which allows to solve arbitrary variational problems, including the hyperelastic problems with arbitrary constituent relations.

\begin{table}[t]
    \caption{Elastic moduli of a composite with the volume fraction $c=0.4\%$ for different number of finite elements $N_k$. Relative moduli of the form $\Delta K/c = (K_{\rm eff} - K_0)/c$ are given in GPa units. Extrapolated values are denoted by the infinity symbol. The last line of the table was calculated using the final analytical expression (\ref{eq:final}). }
    \label{tab:comp}
    \centering
    \begin{tabular}{@{\hskip3pt}Sc@{\hskip10pt}Sc@{\hskip10pt}Sc@{\hskip10pt}Sc@{\hskip10pt}Sc@{\hskip10pt}Sc@{\hskip10pt}Sc@{\hskip3pt}}
        \hline
        $k$&$N_k$&$\Delta K/c$&$\Delta\mu/c$&$\Delta l/c$&$\Delta m/c$&$\Delta n/c$\\
        \hline
        0        & 4736          & 0.25682 & 0.50149 & 28.066 & $-3.180$ & 3.342 \\
        1        & 37888         & 0.25647 & 0.49276 & 28.050 & $-3.152$ & 2.452 \\
        2        & 303104        & 0.25638 & 0.49048 & 28.046 & $-3.140$ & 2.223 \\
        $\infty$ & extrapolation & 0.25635 & 0.48972 & 28.045 & $-3.137$ & 2.146 \\
        --       & theory        & 0.25635 & 0.48971 & 28.045 & $-3.135$ & 2.145 \\
        \hline
    \end{tabular}
\end{table}

The final equation (\ref{eq:final}) is precise for a small volume fraction of inclusions. In order to check this result, we consider a composite with a small inclusion volume fraction $c = 0.4\%$, which corresponds to the inclusion radius $R\approx L/10$ (Fig.~\ref{fig:fem}). We calculate effective moduli using several meshes with different mesh density. The basic mesh consists of $N_0=4736$ finite elements and is presented in Fig.~\ref{fig:fem}. Then we iteratively double the number of finite elements in each direction to obtain two refined meshes with $N_k = 8^k N_0$ finite elements for $k=1,2$. For each number of finite elements $N_k$, we obtain effective moduli $K_{\rm eff}^{(k)}, \mu_{\rm eff}^{(k)}, l_{\rm eff}^{(k)}, m_{\rm eff}^{(k)}, n_{\rm eff}^{(k)}$. The precision of the obtained moduli is scaled quadratically with the size of finite elements: $K_{\rm eff}^{(1)} - K_{\rm eff}^{(0)} \approx 4 (K_{\rm eff}^{(2)} - K_{\rm eff}^{(1)})$ and the same for other moduli. As a result, we can extrapolate moduli for $N_k\to\infty$ as $K_{\rm eff}^{(\infty)} =  (4 K_{\rm eff}^{(2)} - K_{\rm eff}^{(1)})/3$.  The results are shown in Table~\ref{tab:comp} as relative moduli $\Delta K/c = (K_{\rm eff} - K_0)/c$, $\Delta \mu/c = (\mu_{\rm eff} - \mu_0)/c$, etc, which mean the influence of inclusions on the effective moduli of the composite. Table~\ref{tab:comp} shows that the extrapolated values of the relative moduli are very close to the theoretical values: all of them coincide with 3--5 decimal places. Therefore, we have validated our main result (\ref{eq:final}) for a small volume fraction of inclusions.

In order to study variable elastic contrast of the composite, we multiply all elastic moduli of the inclusion (linear and nonlinear) by a coefficient $\alpha$. In this case matrix and inclusions have elastic moduli $K_0,\mu_0,l_0,m_0,n_0$ and $\alpha K_1,\alpha \mu_1,\alpha l_1,\alpha m_1,\alpha n_1$ respectively. For a small value of the parameter $\alpha\ll1$, we have relatively soft inclusions. In the opposite case, $\alpha\gg1$, inclusions are rigid. In both limiting cases, elastic moduli of inclusions do not affect the effective moduli of the composite.

\begin{figure}[t]
    \centering
    \includegraphics[scale=0.6]{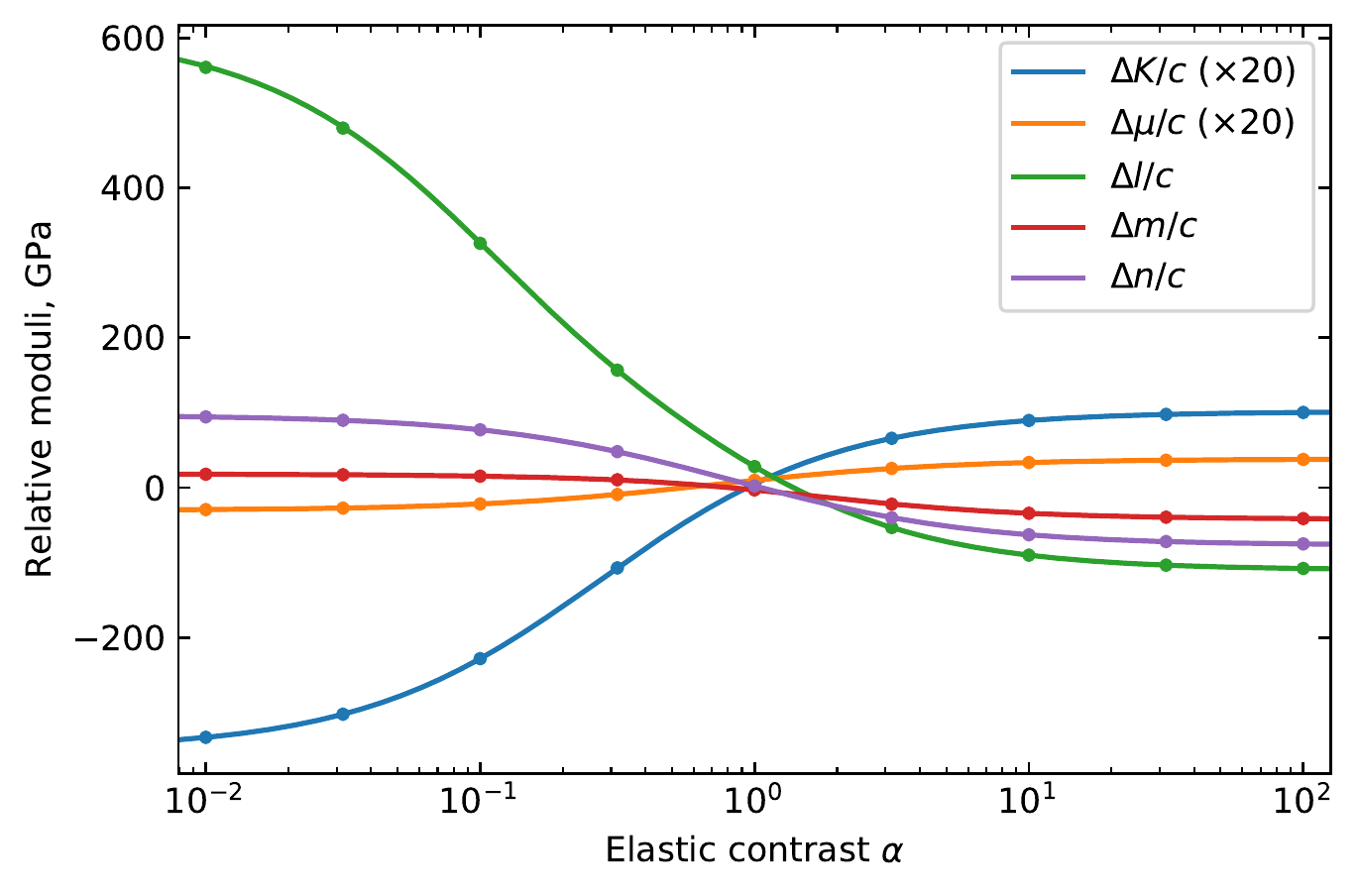}
    \caption{(Color online) Relative moduli of the composite with the volume fraction $c=0.4\%$ for a variable elastic contrast defined by the parameter $\alpha$. Linear relative moduli $\Delta K/c$ and $\Delta \mu/c$ are multiplied by 20 for better visual performance. Lines represent the final equation (\ref{eq:final}). Dots were obtained using the finite element method.}
    \label{fig:alpha}
\end{figure}

For different values of the parameter $\alpha$, we apply the same finite element method with extrapolation procedure for the same volume fraction $c=0.4\%$. Figure \ref{fig:alpha} shows that the final equation (\ref{eq:final}) coincide with the numerical results for any value of the parameter $\alpha$. The point $\alpha=1$ corresponds to the values presented in Table~\ref{tab:comp}. For soft inclusions ($\alpha\ll1$) and rigid inclusions ($\alpha\gg1$), one can see an increased influence of inclusions on effective elastic moduli.

\begin{figure}[t]
    \centering
    \includegraphics[scale=0.6]{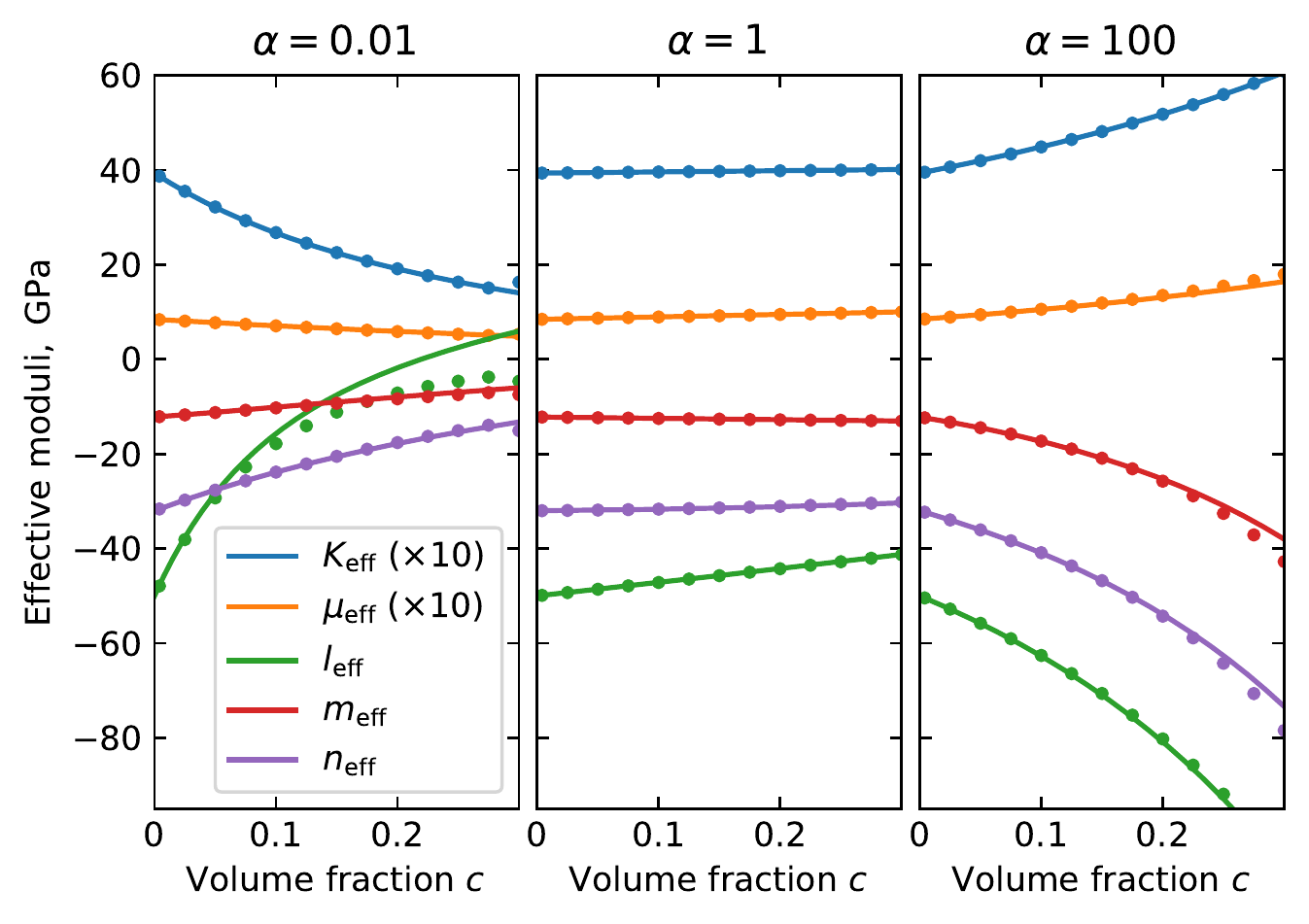}
    \caption{(Color online) Effective moduli of the composite for different volume fraction $c$ for three different elastic contrasts defined by the parameter $\alpha$. Linear effective moduli $K_{\rm eff}$ and $\mu_{\rm eff}$ are multiplied by 10 for better visual performance. Lines represent the final equation (\ref{eq:final}). Dots were obtained using the finite element method.}
    \label{fig:conc}
\end{figure}

The final equation (\ref{eq:final}) can be applied for the finite volume fraction of inclusions $c$. In this equation, we do not take into account a precise form of the interaction between inclusions. Therefore, the final equation (\ref{eq:final}) is approximate for finite values of $c$. In order to check the precision of the obtained effective moduli, we use the same finite element method for a variable inclusion radius $R$ using the mesh with a fixed number of finite elements $N_1=37888$. The effective moduli are shown in Fig.~\ref{fig:conc} for different volume fraction $c$ and different elastic contrasts: $\alpha=0.01$ (soft inclusions), $\alpha=1$ (polystyrene inclusions), and $\alpha=100$ (rigid inclusions). Despite the fact that the interaction between inclusions was not calculated precisely, the theory agrees well with the numerical results. The deviation of the effective moduli rarely exceeds 1\%. The cases $\alpha=0.01$ and $\alpha=100$ demonstrate a noticeable nonlinear dependence on the effective moduli on the volume fraction $c$, which also coincides with the theory. There is only one visible deviation between the theory and the numerical calculation, which is the modulus $l_{\rm eff}$ for soft inclusions ($\alpha=0.01$) for $c\geq0.1$. However, this modulus has the strongest sensitivity on the volume fraction $c$. This deviation can also be caused by the numerical calculation which considers one inclusion in a periodic cell. It effectively acts as an infinite composite with inclusions placed in a simple cubic lattice. The precise calculation of effective moduli of a large volume with randomly located inclusions requires much more computational resources. Accounting of the mutual arrangement of inclusions is the subject of future studies for both theory and numerical calculations.

\section{Conclusions}
\label{sec:conclusion}

In this paper we have derived analytical equations governing the effective nonlinear elastic moduli of a composite with a nonlinear matrix and nonlinear elastic spherical inclusions. The effective nonlinear moduli $l_{\rm eff},m_{\rm eff},n_{\rm eff}$ depend linearly on the nonlinear moduli of the constituents $l_0,m_0,n_0,l_1,m_1,n_1$. At the same time, the dependence on the linear moduli $K_0,\mu_0,K_1,\mu_1$ has a complicated nonlinear form. We have found the ``geometric'' contribution $l_g,m_g,n_g$ to the effective nonlinear moduli, which is due to the inhomogeneous structure of the composite and is defined solely by the linear moduli. The proposed methodology can be generalized to a case of more complex inclusions, such as e.g. those of ellipsoidal shape.

The obtained expressions have been verified using the finite element method for various elastic contrast of inclusions. For a small volume fraction of inclusions $c$, the obtained results were verified with 3--5 significant digits. For finite values of $c$, the analytical expressions agree well with the numerical data and the deviation of the effective moduli rarely exceeds 1\%.

Despite a rather complicated analytical form, final equations can be easily used for computation of effective elastic moduli for given elastic moduli of the matrix and inclusions. Various examples of this procedure can be found on GitHub (\cite{nonlinearmoduli}) in the form of Wolfram Mathematica function, Python code, or an interactive XLS table.  

The financial support from Russian Science Foundation (RSF), under the grant no.~17-72-20201, is gratefully acknowledged.

\appendix

\section{Coefficients of the final equation}
\label{app}
In this appendix we provide the matrices $\mathbf{P}^{0}$ and $\mathbf{P}^{1}$ and geometric nonlinearity modules $l_g$, $m_g$, $n_g$, which are included in the final equation (\ref{eq:final}). In order to shorten the expressions, we introduce the following notations:
\begingroup
\allowdisplaybreaks
\begin{flalign}
    &a = 3 K_0 + 4 \mu_0, &\\
    &b = 3 K_1 + 4 \mu_0, &\\
    &d = \mu_0 - \mu_1, &\\  
    &e = \mu_0((6c+9)K_0+4(3c+2)\mu_0)+6(1-c)\mu_1\left(K_0+2\mu_0\right), &\\
    &f = 3c(K_0 - K_1) + 3 K_1+4 \mu_0. &
\end{flalign}
Note that $e$ and $f$ are denominators in the linear effective moduli (\ref{eq:mu_eff}) and (\ref{eq:K_eff}) respectively. Using the above notation, elements of the matrices $\mathbf{P}^{0}$ and $\mathbf{P}^{1}$ can be written in the following form
\begin{flalign}
    &P_{11}^0 = -\frac{P_{31}^0}{9} - \frac{(3a-2b)b^2+3(a-b)^2bc+(a-b)^3c^2}{f^3}, &\\
    &P_{12}^0 = -\frac{P_{32}^0}{9} + \frac{2(1-c)(a-b)^2b}{f^3}, &\\
    &P_{13}^0 = -\frac{P_{33}^0}{9} + \frac{c(1-c)(a-b)^3-a^3}{9f^3}, &\\
    &P_{21}^0 = (1-c)\frac{120\mu_0^2d^2}{e^2}\left(1-\frac{10d\mu_0}{7e}-\frac{c(a-b)}{f}\right), &\\
    &P_{22}^0 = \frac{P_{32}^0}{6} - P_{22}^1 - 6P_{23}^0 - 6P_{23}^1 + P_{33}^0 + P_{33}^1 \notag &\\*
        &\quad+(1-c)\frac{d}{fe^2}\left(4\mu_0^2d(39b-19a)+a(a-b)\left(a+4\mu_0\right)\left(11\mu_0+9\mu_1\right)\right) &\\
    &P_{23}^0 = \frac{P_{33}^0 + P_{33}^1}{6} - P_{23}^1 -(1-c)\frac{d^2}{6fe^2}\left(5a^2(a+b)+4c(a-b)(a+2\mu_0)^2\right), &\\
    &P_{31}^0 = -(1-c)\frac{7200 \mu_0^3 d^3}{7 e^3}, &\\
    &P_{32}^0 = (1-c)\frac{180 \mu_0d^2}{7 e^3} \left(a(a+4\mu_0)(11\mu_0+9\mu_1)-76\mu _0^2d\right),&\\
    &P_{33}^0 = -P_{33}^1+(1-c)\frac{d^2}{7e^3}\left(56c\left(a+2\mu_0\right)^3d-75a^2\left(\mu_0d-a\left(5\mu_0+2\mu_1\right)\right)\right), &\\
    &P_{11}^1 = \frac{a^3}{f^3}, &\\
    &P_{12}^1 = P_{21}^1 = P_{31}^1 = P_{32}^1 = 0, &\\
    &P_{13}^1 = \frac{a^3}{9} \left(\frac{1}{f^3} - \frac{125 \mu _0^3}{e^3}\right), &\\
    &P_{22}^1 = \frac{25 a^3 \mu _0^2}{f e^2}, &\\
    &P_{23}^1 = (1-c)\frac{25a^3\mu_0^2}{6fe^3}\bigl(5b\mu_0+4\mu_0d-a\left(3\mu_0+2\mu_1\right)\bigr), &\\
    &P_{33}^1 = \frac{125 a^3 \mu_0^3}{e^3}. &
\end{flalign}
The geometrical contribution to nonlinear moduli has the following form: 
\begin{flalign}
    &l_g =  -\frac{n_g}{9} + (1-c)\frac{(a-b)^2}{6f^3}\Bigl(3ab-6(a+b)\mu_0-4c(a-b)\mu_0\Bigr), &\\
    &m_g = \frac{n_g}{6} - (1-c)\frac{2d}{3fe^2}\Bigl(63b\mu_0^3d+10ca\mu_0d(a-b)(a+2\mu_0) &\notag\\*
    &\qquad\qquad + 3a\mu_0^2\bigl(19\mu_0d-4b(\mu_0-6\mu_1)\bigr)+a^3d(13\mu_0-5b) &\notag\\*
    &\qquad\qquad\qquad\qquad\qquad\qquad + a^2\mu_0\bigl(7bd-6\mu_0(3\mu_0+7\mu_1)\bigr)\Bigr), &\\
    &n_g = -(1-c)\frac{12\mu_0d^2}{7e^3}\Bigl(7cd(3a-4\mu_0)(a+2\mu_0)^2 &\notag\\*
    &\qquad\qquad\qquad -5\bigl(6\mu_0^3d+10a^3\mu_1-a^2\mu_0d+12a\mu_0^2(3\mu_0+7\mu_1)\bigr)\Bigr).&
\end{flalign} 
\endgroup
 
\bibliographystyle{elsarticle-harv} 
\bibliography{base}

\end{document}